\documentstyle[12pt]{article}
\begin{document}
\begin{titlepage}
\begin{center}
\hspace*{6cm} Preprint IFUNAM  FT-93-017\\
\hspace*{6cm} May 1993 \\
\hspace*{6cm} (Russian version: December 1992)\\
\vspace*{8mm}
{\large{\bf DECAY RATE OF A POSITRONIUM. \\ \vskip1mm
REVIEW OF THEORY AND EXPERIMENT}}\\
\vspace*{4mm}
{\bf VALERI V. DVOEGLAZOV}$^{*,\,\dagger}$\\
\vskip1mm
{\it Departamento de F\'{\i}sica Te\'orica\\
Instituto de F\'{\i}sica, UNAM, Apartado Postal 20-364\\
 01000 Mexico, D. F.\,\,  MEXICO}\\
\vskip2mm and\\
{\bf RUDOLF N. FAUSTOV} \\
\vskip1mm
{\it  Scientific Council for Cybernetics\\
Russian Academy of Sciences, Vavilov str. , 40\\
Moscow  117333  RUSSIA}\\
\vskip2mm and\\
{\bf YURI N. TYUKHTYAEV}$^{\ddagger}$\\
\vskip1mm
{\it Department of Theoretical} \& {\it Nuclear Physics\\
Saratov State University, Astrakhanskaya str. , 83\\
Saratov  410071  RUSSIA} \\
\end{center}
\vspace*{3mm}
\begin{abstract}
The present status of theoretical and experimental investigations of
the decay rate of a positronium is considered. The increasing
interest
to this problem  has been caused by the disagreement of the
calculated value
of $\Gamma_3 (o-Ps)$ and the recent series of precise experiments.
The
necessity of new calculations on the basis of the quantum field
methods in bound state theory is pointed out  with taking into
account the
dependence
of the interaction kernel on relative energies.
\end{abstract}

\vspace*{6mm}
\centerline{{\bf Submitted to "Modern Physics Letters A"}}
\vspace*{7mm}


\vspace*{-5mm}
\noindent
----------------------------------------------------------------\\
\vspace*{-5mm}

$^{*}$ On leave from: {\it Dept.Theor.} \& {\it Nucl. Phys.,
   Saratov State University\\and Sci.} \& {\it Tech. Center for
Control
and Use of Physical Fields and Radiations\\Astrakhanskaya str. ,
83,\,\,
   Saratov 410071 RUSSIA}

$^{\dagger}$ Email: valeri@ifunam.ifisicacu.unam.mx\\
\hspace*{22mm}dvoeglazov@main1.jinr.dubna.su

$^{\ddagger}$ Email: postmaster@ccssu.saratov.su
\end{titlepage}

\setcounter{page}{1}
The quantum-electrodynamic systems, consisting of particle and
anti-particle,
have specific features. Apart from a scattering channel, the
annihilation
channel appears in this case. A positronium atom, which is a specimen
of these
systems, has no stability. The life time of a positronium (or the
decay rate)
is the subject of precise experimental and theoretical
investigations. The
charge parity of a positronium, $C = (-1)^{L+S}$
($L $ is the eigenvalue of an angular momentum operator, $S$ is the
eigenvalue
of a total spin operator for a system under consideration), is a
motion
constant. Consequently, all its states are separated into the charge
- even
states ($S=1$) and the charge - odd ones ($S=-1$). The positronium
total spin
is also conserved and the energy levels are classified as the singlet
levels
($S=0$,  parapositronium)  and as the triplet ones ($S=1$,
orthopositronium).
The S -- state ($L=0$)  parapositronium has a positive parity and the
S --
state orthopositronium has a negative parity. As a consequence of
conservation
of a charge parity in electromagnetic interaction a parapositronium
is
disintegrated into the even number of photons and an orthopositronium
is
decayed into the odd one.

At the present time, the essential disagreement is turned out between
the
theoretical and experimental values for the decay rate of an
orthopositronium.
The theoretical predictions are ~\cite{pg1}-\cite{pg4}:\\
\begin{eqnarray}
\lefteqn{\Gamma^{theor}_3(o-Ps)=\frac{\alpha^6
mc^2}{\hbar}\frac{2(\pi^2-9)}{9\pi} \left
[1+A_3\frac{\alpha}{\pi}-\frac{1}{3}\alpha^2
ln\alpha^{-1}+B_3(\frac{\alpha}{\pi})^2+\ldots \right ] = }
\nonumber\\
&=&\Gamma_0+\frac{m\alpha^7}{\pi^2}\left \{ {-1.984(2) \choose
-1.9869(6)}\right\}+\frac{m\alpha^8}{\pi}
ln\alpha^{-1}\left [-\frac{4}{9}\zeta(2)+\frac{2}{3}\right
]+\frac{m\alpha^8}{\pi^3}{\cal X}+\ldots \nonumber\\
&=&\left \{ {7.0386(2) \choose 7.03830(7)} \right \}\,\mu s^{-1},
\end{eqnarray}
where
\begin{eqnarray}
A_3^{\cite{pg3}}&=&-10.266\pm 0.011,\\
A_3^{\cite{pg4}}&=&-10.282\pm 0.003.
\end{eqnarray}
Taking into account the modern value of $\alpha$, the fine structure
constant, the result can be recalculated~\cite{pr11}:
\begin{equation}
\Gamma^{theor}_{3}=7.038\,31(5)\,\mu s^{-1}.
\end{equation}

The last experimental measurings are~\cite{pg21,pg22}\footnote
{See Table I for the preceding experimental results.}:
\begin{eqnarray}
\Gamma^{exp}_{\cite{pg21}}(o-Ps)&=&7.0514(14)\,\mu s^{-1}\\
\Gamma^{exp}_{\cite{pg22}}(o-Ps)&=&7.0482(16)\,\mu s^{-1}.
\end{eqnarray}
The result of Ref.~\cite{pg21}  has  9.4 standard deviation from the
predicted theoretical decay rate and the result of  ~\cite{pg22} has
6.2
standard deviation.
The coefficient $B_3=1$ in the $O(\alpha^8)$ term
can contribute $3.5\cdot10^{-5}\mu s^{-1}$ (or 5 ppm of  $\Gamma_3$)
only.
To take off the above disagreement the coefficient $B_3$ has to be
equal to
about $\simeq 250\pm 40$, what is very unlikely, indeed. However,
this
opportunity is pointed out in ~\cite{pg22} just not to be rejected
{\it a
priori}.  Therefore, the calculation of  $B_3$  coefficient  is
desirable
enough now.

For the first time the main contribution in the orthopositronium
decay rate has
 been calculated in ~\cite{pg1} :
\begin{equation}
\Gamma_0(o-Ps)=-2Im(\Delta
E_{3\gamma})=\frac{2}{9\pi}(\pi^2-9)m\alpha^6=7,211\,17\,\mu s^{-1}.
\end{equation}
The corrections of the $O(\alpha)$  order of the magnitude to this
quantity had
been calculating by numerical method ~\cite{pg2,pg4,pg3a,pg3b} at
first,
but later  some of them has been figure out analytically
in~\cite{pg3,a35}
and~\cite{pr7}-\cite{pr9} in the Feynman gauge.
The corrections, which  come from the diagrams with the self-energy
and vertex
insertions, have been calculated by Adkins ~\cite{pr8,pr9}:
\begin{eqnarray}
\Gamma_{OV}&=&\Gamma_0\frac{\alpha}{\pi} \left \{
           D+\frac{3}{4(\pi^2-9)}\left [ -26-\frac{115}{3}ln2
+\frac{91}{18}\zeta(2)+\frac{443}{54}\zeta(3)+
\frac{3419}{108}\zeta(2)
ln2-\right .\right .\nonumber\\
           &-&\left.\left. R \right ] \right
\}=\Gamma_0\frac{\alpha}{\pi}\left
[ D+2.971\,138\,5(4)\right ],\\
\Gamma_{SE}&=&\Gamma_{0}\frac{\alpha}{\pi} \left \{ -D-4+\frac{3}{ 4
(\pi^2-9)
} \left [ -7+\frac{67}{3}ln
2+\frac{805}{36}\zeta(2)-
\frac{1049}{54}\zeta(3)-\right.\right.\nonumber\\
           &-&\left.\left.\frac{775}{54}\zeta(2)ln 2 \right ] \right
\}=\Gamma_0\frac{\alpha}{\pi}\left [ -D+0.784\,98\right ],\\
\Gamma_{IV}&=&\Gamma_{0}\frac{\alpha}{\pi}\left \{\frac{1}{2}
D+\frac{3}{4
(\pi^{2}-9) } \left [
-4-\frac{34}{2}ln2-\frac{841}{36}\zeta(2)+\frac{1253}{36}\zeta(2)ln
2+\frac{1589}{54}\zeta(3)+\right.\right.\nonumber\\
&+&\left.\left.\frac{17}{40}\zeta^{2}(2)-\frac{7}{8}\zeta(3)ln
2+\frac{5}{2}\zeta(2)ln^{2} 2-\frac{1}{24} ln^{4} 2-a_{4} \right ]
\right
\}=\nonumber\\
           &=&\Gamma_{0}\frac{\alpha}{\pi}\left
[\frac{1}{2}D+0.160\,677\right
],
\end{eqnarray}
where
\begin{equation}
R=\int \limits^{1}_{0} dx\frac{ln(1-x)}{2-x}\left
[\zeta(2)-Li_2(1-2x)\right
]=-1.743\,033\,833\,7(3),\\
\end{equation}
\begin{equation}
a_4=Li_4(\frac{1}{2})=\sum_{n=1}^{\infty}\frac{1}{n^4
2^n}=0.517\,479\,061\,674,
\end{equation}
\begin{equation}
\zeta(2)=\frac{\pi^2}{6},\quad \zeta(3)=1.202\,056\,903\,2,
\end{equation}
and
\begin{equation}
D=\frac{1}{2-w}-\gamma_E+ln(4\pi)
\end{equation}
is   standard expression of a dimensional regularization ($2\omega$
is a space
dimension).  The indices "$IV$" and "$OV$" designate the insertions
in the
internal photon-electron vertex and in the outer ones,
correspondingly.
The above results are  co-ordinated with the Stroscio's
result~\cite{pr7} when
\begin{equation}
\Gamma_0\frac{\alpha}{\pi}\left [-D-4-2 ln (\lambda^2/m^2)\right ]
\end{equation}
being added to the last one. This is necessary because of the
different
regularization procedures which have been used in~\cite{pr7} and
in~\cite{pr8,pr9}, correspondingly.

Recently, the calculations of these corrections  have been finished
{}~\cite{pr11} in the Fried -- Yennie gauge:
\begin{eqnarray}
\Gamma_{SE}&=&\frac{m\alpha^7}{\pi^2}\left
[-\frac{13}{54}\zeta(3)+\frac{461}{108}\zeta(2)ln2
-\frac{251}{72}\zeta(2)-\frac{29}{6}ln2+\frac{9}{2}\right
]=\nonumber\\
&=&\frac{m\alpha^7}{\pi^2}(-0.007\,132\,904)=\Gamma_0\frac{\alpha}{\pi
}(-0.036\,911\,113),\\
\Gamma_{OV}&=&\frac{m\alpha^7}{\pi^2}\left
[-\frac{88}{54}\zeta(3)-\frac{299}{216}\zeta(2)ln2
+\frac{49}{18}\zeta(2)+\frac{13}{6}ln2-2-\frac{1}{6}R\right
]=\nonumber\\
&=&\frac{m\alpha^7}{\pi^2}(0.732\,986\,380)=\Gamma_0\frac{\alpha}{\pi}
(3.793\,033\,599).
\end{eqnarray}
The contributions from the remained diagrams  ( with a radiative
insertion in a
vertex of an internal photon;  with two radiative photons spanned;
the diagram
taking into account boundary effects and the annihilation diagram,
see Fig. I
in [5b] ),
have been calculated numerically. Totally, the $O(\alpha)$
corrections are
joined to give
\begin{equation}
\frac{m\alpha^7}{\pi^2}\left [-1.987\,84(11)\right ]=
\Gamma_0\frac{\alpha}{\pi}\left [-10.286\,6(6)\right ].
\end{equation}
 Then we have\footnote{The uncalculated yet $O(\alpha^8)$ corrections
are not
accounted here.} :
\begin{equation}
\Gamma_{3,~\cite{pr11}}^{theor}(o-Ps)=7.038\,236(10)\, \mu s^{-1}.
\end{equation}
The above result is the most precise theoretical result available at
the
present moment.

To solve the existing disagreement between theory and experiment,
the 5 --
photon mode of  $o-Ps$ decay and the 4 -- photon mode of  $p-Ps$
decay have
been under consideration in ~\cite{pru1,pru2}\footnote{ As a
consequence of a
conservation of an angular momentum and an isotropic properties of
coordinate
space an orthopositronium has to decay into the odd number of photons
and a
parapositronium has to decay into the even one,  as outlined
before.}.  The
following theoretical evaluations have been obtained:
\begin{eqnarray}
\frac{\Gamma_5^{\cite{pru1}}(o-Ps)}{\Gamma_3(o-Ps)}&=&0.177
(\frac{\alpha}{\pi})^2\simeq 0.96\cdot10^{-6},\\
\frac{\Gamma_4^{\cite{pru1}}(p-Ps)}{\Gamma_2(p-Ps)}&=&0.274
(\frac{\alpha}{\pi})^2\simeq 1.48\cdot10^{-6},
\end{eqnarray}
and
\begin{eqnarray}
\Gamma_5^{\cite{pru2}}(o-Ps)&=&0.018\,9(11)\alpha^2\Gamma_0,\\
\Gamma_4^{\cite{pru2}}(p-Ps)&=&0.013\,89(6)m\alpha^7.
\end{eqnarray}
They are in agreement with each other and with the results of the
previous
papers~\cite{pru3}\footnote{The result~\cite{pru4} is not correct,
four times
less than the above cited results. The explanation of this was given
in~\cite{pru2}.} :
\begin{equation}
\Gamma_4^{~\cite{pru3}}(p-Ps) = 0.013\,52\,m\alpha^7 = 11.57\cdot
10^{-3}
s^{-1}.
\end{equation}

In the connection with the present situation with respect to the
decay rate of
$o-Ps$  investigations of alternative decay modes for this system
(e.g. $o-Ps\rightarrow\gamma+a$, $a$ is an axion, a pseudo-scalar
particle with
mass  $m_a<2m_e$) are of present interest ~\cite{pru6}-\cite{pru6d}.
In the
article ~\cite{pru6c} the following experimental limits of the
branching of
decay width
have been obtained:
\begin{equation}
Br=\frac{\Gamma(o-Ps\rightarrow \gamma+a)}{\Gamma(o-Ps\rightarrow
3\gamma)}
< 5\cdot 10^{-6} - 1\cdot 10^{-6}\quad (30\, ppm),
\end{equation}
provided that  $m_a$ is in the range 100 -- 900 keV.  In the case of
the axion mass less than 100 keV (which is implied by the Samuel's
hypothesis
{}~\cite{pru6ca}; according to the cited paper~\footnote{The proposed
values do
not influence the agreement  of theoretical and experimental results
of an
anomalous magnetic moment (AMM) of an electron.}\,
$m_a< 5.7 \,keV$, $g_{ae^{+} e^{-}} \sim 2\cdot 10^{-8}$) the limits
of
$Br$ are the following ones~\cite{pru6d} :
\begin{eqnarray}
Br&=&7.6\cdot 10^{-6},\quad \mbox{if}\,\,\,m_a \sim 100\, keV,\\
Br&=&6.4\cdot 10^{-5}, \quad \mbox{if}\,\,\, m_a < 30\, keV.
\end{eqnarray}
These limits are about 2 orders less than the value which is
necessary to
remove the disagreement.

Finally, a decay $o-Ps \rightarrow nothing$ (that is  into
weak-interacting
non-detected particles)~\footnote{Analogously to   Glashow's
hypothesis of the
decay into invisible "mirror" particle~\cite{pru6cb}.} has been
investigated in
{}~\cite{pru6e}.
The obtained result
\begin{equation}
\frac{\Gamma(o-Ps\rightarrow nothing)}
{\Gamma(o-Ps\rightarrow 3\gamma)}< 5.8\cdot 10^{-4} \quad (350\, ppm)
\end{equation}
expels the opportunity that this decay mode is an origin of
disajustment
between theory and experiment.

The decay of $o-Ps$ into two photons, which breaks the CP --
invariance, as
else mentioned in ~\cite{pru7a,pru7b}, was experimentally rejected
in~\cite{pru7}
\footnote{The physics ground of these speculations is a possible
existence of
an unisotropic vector field with non-zero  vacuum expectation
{}~\cite{pru7c},
with whom an electron and a positron could be interacting
\begin{equation}
{\cal L}=g\bar\psi O_{\alpha\beta}\psi A^{\alpha}\Omega^{\beta},
\end{equation}
$\cal L$ is the interaction Lagrangian.}.

Let us mention,  the contribution of weak interaction  has been
studied in
{}~\cite{pw1}.
However, because of the factor  $m_e^2/M^2_W \sim G_F \cdot m_e
\simeq 3 \cdot
10^{-12}$  it cannot influence final results. In the cited article
the weak
decay modes are estimated as
\begin{equation}
\frac{\Gamma(p-Ps\rightarrow 3\gamma)}{\Gamma(p-Ps\rightarrow
2\gamma)}
\simeq\frac{\Gamma(o-Ps\rightarrow 4\gamma)}{\Gamma(o-Ps\rightarrow
3\gamma)}
\simeq\alpha(G_F m_e^2 g_V)^2\simeq 10^{-27},
\end{equation}
where $G_F$ is the Fermi constant for weak interaction,
\begin{equation}
g_V=1-4sin^2\Theta_W\simeq 0.08,
\end{equation}
$\Theta_W$  is the Weinberg angle.
The current experimental limits are ~\cite{pw2,pw3} :
\begin{eqnarray}
\frac{\Gamma(p-Ps\rightarrow 3\gamma)}{\Gamma(p-Ps\rightarrow
2\gamma)}&\leq&2.8\cdot 10^{-6},\\
\frac{\Gamma(o-Ps\rightarrow 4\gamma)}{\Gamma(o-Ps\rightarrow
3\gamma)}&\leq&8\cdot 10^{-6}.
\end{eqnarray}

In the Table I  all experimental results for the $o-Ps$ decay rate ,
known to
us, are presented \footnote{The results of the papers
{}~\cite{pg11aa,pg11a} and
{}~\cite{pg12a}
can be accounted as rough estimations only.}.\\
\begin{center}
Table  I.  The experimental results for the $o-Ps$  decay rate.\\
\vspace*{5mm}
\begin{tabular}{||c|c|l|r|c||}
\hline
\hline
Year &Reference&$\Gamma_3(o-Ps),\,\mu s^{-1}$&Error,\,ppm&Technique\\
\hline
1968&~\cite{pg11b}&7.262(15)&2070&gas\\
1973&~\cite{pg11}&7.262(15)&2070&gas\\
1973&~\cite{pg12}&7.275(15)&2060&gas\\
1976&~\cite{pg13}&7.104(6)&840&powder $SiO_{2}$\\
1976&~\cite{pg14}&7.09(2)&2820&vacuum\\
1978&~\cite{pg15}&7.056(7)&990&gas\\
1978&~\cite{pg16}&7.045(6)&850&gas\\
1978&~\cite{pg17}&7.050(13)&1840&vacuum\\
1978&~\cite{pg17a}&7.122(12)&1680&vacuum\\
1982&~\cite{pg18}&7.051(5)&710&gas\\
1987&~\cite{pg19}&7.031(7)&1000&vacuum\\
1987&~\cite{pg20}&7.0516(13)&180&gas\\
1989&~\cite{pg21}&7.0514(14)&200&gas\\
1990&~\cite{pg22}&7.0482(16)&230&vacuum\\
\hline
\hline
\end{tabular}
\end{center}
\vspace*{5mm}

Regarding the results for the decay rate of a parapositronium, the
situation
was highly favorable until the last years. The theoretical value,
which was
found out else in the fifties ~\cite{pg52r,pg52}, is equal to
\begin{equation}
\Gamma_2^{theor}(p-Ps)=-2Im(\Delta
E_{2\gamma})=\frac{1}{2}\frac{\alpha^5
mc^2}{h}\left [ 1-\frac{\alpha}{\pi}(5-\frac{\pi^2}{4})\right ]=
7.9852\,
ns^{-1}.
\end{equation}
The above  value, confirmed in  ~\cite{pg53,pg54},  coincides with
the direct
experimental result up to the good accuracy:
\begin{equation}
\Gamma^{exp}_{\cite{pg18}}(p-Ps)=7.994 \pm 0.011\, ns^{-1}.
\end{equation}

The experimental values of the parapositronium decay rate are showed
in the
Table II \footnote{The branching of the decay rates of a para-
and an orthopositronium $\frac{\Gamma_2(p-Ps)}{\Gamma_3(o-Ps)}$ had
been
measuring in the experiments of 1952 and 1954 only.
The presented results  are recalculated by means of the first direct
experimental value   $\Gamma_3(o-Ps)=7.262(15)\,\mu s^{-1}$
\,\,~\cite{pg11b}.}.\\
\begin{center}
Table II.  The experimental results for the $p-Ps$ decay rate.\\
\vspace*{5mm}
\begin{tabular}{||c|c|l|r|c||}
\hline
\hline
Year&Reference&$\Gamma_2(p-Ps),\,ns^{-1}$&Error, $\%$&Technique\\
\hline
1952&~\cite{p2wh}&7.63(1.02)&13&gas\\
1954&~\cite{p5a}&9.45(1.41)&15&gas\\
1970&~\cite{pg55}&7.99(11)&1.38&gas\\
1982&~\cite{pg18}&7.994(11)&0.14&gas\\
\hline
\hline
\end{tabular}
\end{center}
\vspace*{5mm}

In the articles~\cite{pg3,pg54} it was pointed out that it is
necessary to add
the logarithmic corrections on $\alpha$ to the Harris and Brown's
result.   In
the article [13a]
these corrections to the $\Gamma_3(o-Ps)$ and $\Gamma_2(p-Ps)$ have
been
calculated again, with the result of the decay rate
of a parapositronium  differing with the one found out
before~\cite{pg3,pg54}:
\begin{eqnarray}
\Gamma_{2}^{\cite{a35}}(p-Ps, \alpha^2
ln\alpha)&=&\frac{m\alpha^5}{2}\cdot
2\alpha^2 ln\alpha^{-1},\\
\Gamma_{2}^{\cite{pg3,pg54}}(p-Ps, \alpha^2
ln\alpha)&=&\frac{m\alpha^5}{2}\cdot\frac{2}{3}\alpha^2
ln\alpha^{-1}.
\end{eqnarray}

Finally,  the quite unexpected (and undesirable) result, presented in
the
Remiddi's (and collaborators) talk ~\cite{pg55a} should be mentioned.
The
calculations, carried out by the authors of cited paper, lead to the
additional
contribution:
\begin{equation}
\Gamma_{2}^{\cite{pg55a}}(p-Ps, \alpha ln\alpha)=
\frac{m\alpha^5}{2}(\frac{\alpha}{\pi})2 ln \alpha,
\end{equation}
which is explained by authors to appear as a result of taking into
account the
dependence of an interaction kernel on the relative
momenta.\footnote{Let us
mark, the additional contributions, which are similar to the ones of
Ref.~\cite{pg55a}, appeared in the calculations of the hyperfine
splitting
(HFS) of the ground state of two-fermion system by the quasipotential
method~\cite{Log,Kad}  when  taking into account the dependence of
the
interaction kernel on the relative energies ~\cite{Tyuk, Boi}. For
example, the
$O(\alpha^2 ln\alpha)$ correction to the HFS, obtained from
one-photon diagram,
is equal to
\begin{equation}
\Delta E^{HFS}_{tr}(\alpha^2 ln\alpha)=E_F\frac{\mu^2\alpha^2}{m_1
m_2}
(\frac{m_1}{m_2}+\frac{m_2}{m_1}+2)ln\alpha^{-1},
\end{equation}
when using the version of the quasipotential approach based on the
amplitude $\tau=(\hat{G}^+_0)^{-1}\widehat{G_0
TG_0}^{+}(\hat{G}^+_0)^{-1}$.
Here  $\hat{G}^{+}$ is the two-time Green's function projected onto
the
positive-energy states, $m_1$ and $m_2$ are the masses of the
constituent
particles, $E_F$ is the Fermi energy. The index "{\it tr}" designates
that the
diagram of one-transversal-photon exchange is under consideration.

Using the kernel constructed from the on-shell physical amplitude  we
can find
out that the last term in the brackets disappears.
However, the total result to the HFS accounting all diagrams is the
same in
both of versions of quasipotential approach.

We can also come across with the  similar situation in calculations
of  the
anomalous corrections of $O(\alpha ln\alpha)$ order, which do not
appear in the
method on the mass shell. However, the version, based on the two-time
Green's
function, shows the contraction of these anomalously large terms,
taking into
account two-photon exchange diagrams~\cite{Boi}.}

The above-mentioned shows us at the necessity of a continuation of
calculations of the decay rates of an orthopositronium as well as a
parapositronium employing more accurate relativistic methods, one of
which  is
the quasipotential approach in quantum field theory~\cite{Log,Kad}
giving the
opportunity taking into account of binding effects.

{\bf Acknowledgements}

The authors express their sincere gratitude to B. A. Arbuzov, E. E.
Boos, D.
Broadhurst,  R. Fell, V. G. Kadyshevsky, A. Sultanayev for the
fruitful
discussions;
to the Saratov's Scientific and Technological Center and the CONACYT
( Mexico )
for financial support.  We strongly appreciate also the technical
assistance of
S. V. Khudyakov, G. Loyola and A. S. Rodin.

One of us (V. D.)
is very grateful to his colleagues in the Laboratory of Theoretical
Physics at
the JINR (Dubna) and in the Departamento de F\'{\i}sica
Te\'{o}rica at the IFUNAM for the creation of  the excellent working
conditions.

\end{document}